\documentclass[aps,prl,reprint,superscriptaddress,nofootinbib,longbibliography]{revtex4-2}

\usepackage{amsmath,amssymb,bm}
\usepackage{graphicx}
\usepackage{dcolumn}
\usepackage{booktabs}
\usepackage{xcolor}
\usepackage[colorlinks=true,linkcolor=blue,citecolor=blue,urlcolor=blue]{hyperref}

\newcommand{\vk}{\mathbf{k}}
\newcommand{\vq}{\mathbf{q}}

\newcommand{\Deltad}{\Delta_d}

\newcommand{\vgt}{v_q^{\mathrm{gate}}}

\newcommand{\ASz}{A_{S_z}}
\newcommand{\An}{A_n}
\newcommand{\Vch}{V_{\mathrm{ch}}}

\begin{document}

\title{Rashba Spin Demons in Two-Dimensional $d$-Wave Altermagnets and their Electrostatic Control}

\author{Muhammad Irfan Sarwar}
\affiliation{Department of Physics, Quaid-i-Azam University, Islamabad 45320, Pakistan}

\author{Mohsin Raza}
\affiliation{Department of Physics, Quaid-i-Azam University, Islamabad 45320, Pakistan}

\author{Kashif Sabeeh}
\email{ksabeeh@qau.edu.pk}
\affiliation{Department of Physics, Quaid-i-Azam University, Islamabad 45320, Pakistan}

\date{\today}

\begin{abstract}
Spin demons in $d$-wave altermagnets are acoustic collective excitations produced by nearly out-of-phase motion of spin-split quasiparticle populations. We identify a new mode, Rashba spin demon mode, in 2D $d$-wave altermagnets
in the presence of Rashba Spin-Orbit Coupling (RSOC).  The pristine spin demon mode in altermagnets, without RSOC, is a charge-dark excitation which is invisible to charge-sensitive probes. We show that Rashba spin demon mode is a spin-dominated acoustic mode that is distinct from the pristine mode and has a  finite charge density spectral weight.   RSOC  rotates the quasiparticle spin texture in momentum space, generating mixed charge-spin coherence factors that give the longitudinal spin pole a charge residue. We also show that a nearby metallic gate provides a complementary electrostatic knob by screening the Coulomb interaction and tuning the pole dispersion and linewidth.  Within a charge-spin random-phase approximation, we identify regimes where the Rashba mode is underdamped, charge visible, and still predominantly longitudinal-spin-like.  Our results establish RSOC and electrostatic screening as complementary controls for tuning the spin and charge character of altermagnetic spin demons and making them accessible to charge probes without destroying their spin-dominated character.
\end{abstract}
\maketitle

\emph{Introduction.---}
Collective modes reveal how interactions organize the internal degrees of freedom of an electron liquid.  In a single-component metal or a two-dimensional electron gas, the long-range Coulomb interaction produces plasmons governed by the density response and dielectric screening \cite{BohmPines1953,EhrenreichCohen1959,Lindhard1954,Stern1967,FetterWalecka1971,GiulianiVignale2005}.  In a multicomponent fluid, however, different populations can oscillate out of phase.  If their charge densities nearly cancel, the collective mode can be acoustic and weakly visible in ordinary charge probes.  Pines introduced this possibility as a ``demon'' mode \cite{Pines1956}, and the observation of Pines' demon in Sr$_2$RuO$_4$ showed that weakly charge-coupled collective modes can nevertheless be experimentally relevant \cite{Husain2023}.  Related long-lived collective modes appear in spin-polarized electron fluids when spin-resolved particle-hole continua are sufficiently separated \cite{Agarwal2014,Kreil2015}.

Altermagnets bring this idea into a symmetry-defined magnetic setting.  They combine compensated collinear magnetic order with momentum-dependent spin splitting \cite{Smejkal2022PRX,Smejkal2022Landscape,Mazin2022PRX}.  This symmetry has motivated work on spin-split transport, anomalous Hall responses, spin-transfer effects, and compensated magnetic order \cite{Smejkal2020SciAdv,GonzalezHernandez2021,Jungwirth2016NatNano,Baltz2018RMP,Zutic2004RMP}, and experiments have reported altermagnetic band splitting, domain structure, and split magnons in candidate materials \cite{Krempasky2024Nature,Reimers2024NatComm,Yang2025NatComm,Fedchenko2024SciAdv,Amin2024Nature,Liu2024PRL}.  Recent work showed that $d$-wave altermagnets naturally host spin demons: nearly out-of-phase oscillations of spin-split quasiparticle populations whose damping is reduced because the mode lies in a low-damping interval between spin-resolved continua \cite{Gunnink2025PRL,Rostami2025FermiLiquid}.  The same out-of-phase motion that protects the mode also suppresses its net charge fluctuation, making the pristine spin demon difficult to detect through charge density-sensitive probes.

This Letter addresses the following questions: How is the pristine spin demon mode modified in the presence of RSOC and electrostatic gate? Can one brighten the charge dark pristine spin demon mode to make it charge visible without losing its spin identity?  And can one  externally tune the Rashba spin demon mode and change its spin and charge character?  

\begin{figure}[!htbp]
\centering
\includegraphics[width=0.82\columnwidth]{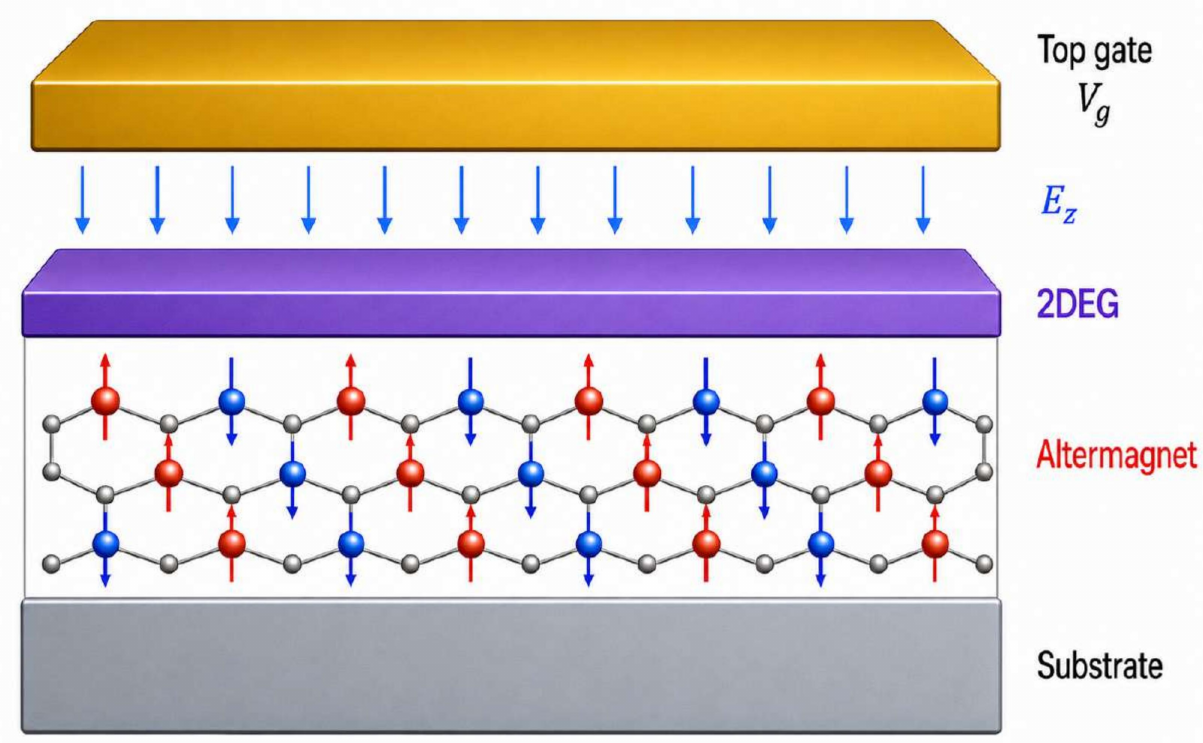}
\vspace{-0.45em}
\caption{Substrate-engineered Rashba altermagnetic platform with electrostatic gate control. A tailored inversion-asymmetric substrate induces Rashba spin-orbit coupling in the two-dimensional conducting layer, while the top gate controls the perpendicular electric field $E_z$ and screens long-range charge fluctuations. The adjacent $d$-wave altermagnet supplies compensated magnetic order and momentum-dependent spin splitting.}
\label{fig:model}
\end{figure}

To answer the first question, we show that Rashba spin demon mode is a spin-dominated acoustic mode that is distinct from the pristine mode and has a  finite charge density spectral weight. In response to the the rest of the questions, a useful control mechanism must satisfy two conditions.  It must create enough spin-charge mixing to give the mode a finite density footprint, but it must not hybridize the acoustic spin demon into an ordinary charge plasmon.  It should also tune the collective denominator, and therefore the mode velocity and linewidth, without relying only on microscopic changes of the altermagnetic band splitting.  Rashba spin-orbit coupling and electrostatic gate screening satisfy these two requirements in complementary ways.   Rashba coupling can be generated by inversion-breaking interfaces, substrates, asymmetric confinement, or a perpendicular electric field \cite{BychkovRashba1984,Nitta1997,Manchon2015,Bihlmayer2022}, changes pure-spin eigenstates into momentum-dependent spinors. Here we consider Rashba spin-orbit coupling generated by tailored substrates. The same spin-orbit physics underlies Edelstein and inverse Edelstein effects, spin-galvanic responses, and spin Hall transport \cite{Edelstein1990,Ganichev2002Nature,Sinova2004PRL,Shen2014PRL,Bercioux2015,Johansson2021PRResearch}.  A nearby metallic gate, by contrast, screens the Coulomb interaction through image charges and changes the collective feedback in the RPA denominator \cite{Chaplik1972,SarmaMadhukar1981,Torre2019,Zabolotnykh2019}.  The platform is shown in Fig.~\ref{fig:model}.  We show that Rashba coupling gives the spin-demon pole a density residue, while the gate tunes its dispersion and quality factor.  The mode is distinct from the pristine spin demon mode and remains dominated by longitudinal spin.

\emph{Single-particle model and spin texture.---}
We model the low-energy quasiparticles of a two-dimensional
$d$-wave altermagnet by
\begin{equation}
H(\vk)=\epsilon_0(\vk)\mathbb{I}
+\Deltad(\vk)\sigma_z
+\alpha_R(k_y\sigma_x-k_x\sigma_y),
\label{eq:Hmain}
\end{equation}
where
\begin{equation}
\epsilon_0(\vk)=\frac{\hbar^2 k^2}{2m_0},
\qquad
\Deltad(\vk)=\frac{\hbar^2}{2m^\ast}(k_x^2-k_y^2).
\label{eq:dwave}
\end{equation}
The term $\Deltad(\vk)\propto k^2\cos2\phi$ carries the
$d$-wave altermagnetic symmetry: it changes sign under a
$\pi/2$ rotation and vanishes along the nodal directions
$k_x=\pm k_y$.  In the spin-conserving limit, this produces
spin-resolved Fermi contours elongated along orthogonal axes.
A density perturbation with fixed propagation direction then
projects differently onto the two spin populations, creating
separated particle-hole continua.  This kinematic separation is
the window in which the pristine altermagnetic spin demon is
formed.

\begin{figure}[!htbp]
\centering
\includegraphics[width=\columnwidth]{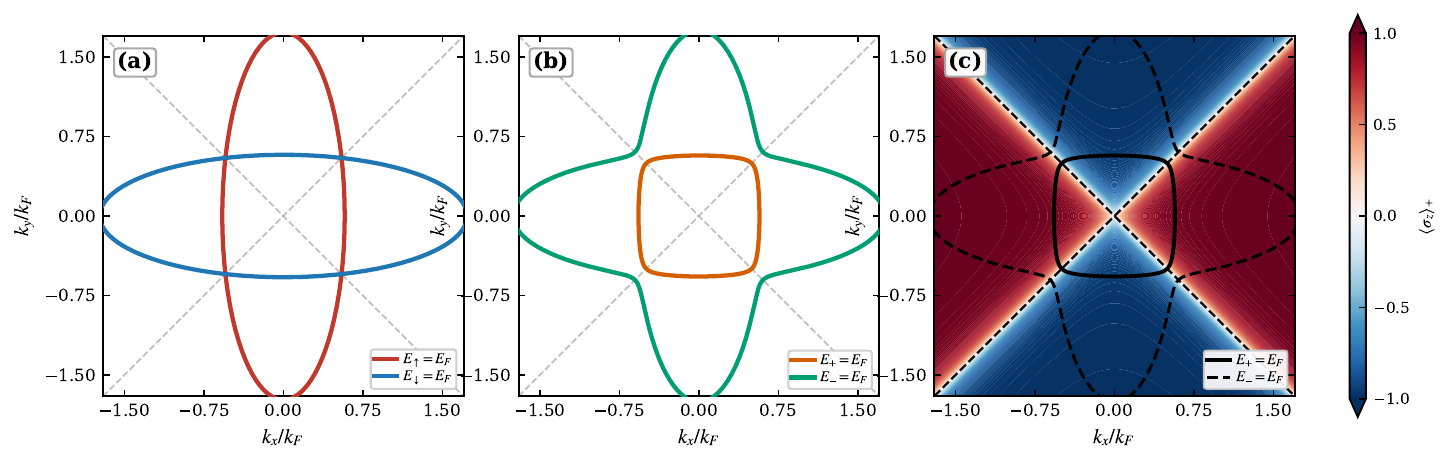}
\vspace{-0.45em}
\caption{Band geometry and Rashba spin texture.  (a) Pristine spin-resolved Fermi contours of the $d$-wave altermagnet.  (b) Rashba-modified Fermi contours at $\bar\alpha_R=0.25$.  (c) Out-of-plane spin texture $\langle\sigma_z\rangle_+$ of the upper Rashba band.  The sign reversal across $k_x=\pm k_y$ reflects the $d$-wave altermagnetic splitting, while the suppression of $|\langle\sigma_z\rangle_+|$ near the nodal directions reflects Rashba spinor mixing.}
\label{fig:band_spin_texture}
\end{figure}

Rashba coupling changes this reference problem in a more
subtle way than a simple band shift.  Writing
\begin{equation}
\begin{split}
H(\vk) &= \epsilon_0(\vk)\mathbb{I} + \mathbf d(\vk)\cdot\boldsymbol\sigma,\\
\mathbf d(\vk) &= \bigl(\alpha_R k_y,-\alpha_R k_x,\Deltad(\vk)\bigr),
\end{split}
\label{eq:dvector_main}
\end{equation}
the two Rashba-altermagnetic bands are
\begin{equation}
E_\lambda(\vk)=
\epsilon_0(\vk)+\lambda |\mathbf d(\vk)|,
\qquad
\lambda=\pm1 .
\label{eq:bands_main}
\end{equation}
The spin texture is
\begin{equation}
\langle\boldsymbol\sigma\rangle_\lambda
=
\lambda \hat{\mathbf d}(\vk).
\label{eq:spintexturemain}
\end{equation}
Thus Rashba coupling simultaneously reshapes the Fermi
contours and rotates the quasiparticle spin texture away from
the fixed $\sigma_z$ axis.  The first effect changes the
available particle-hole phase space, while the second changes
the charge-spin coherence factors entering the response
function.  This spinor rotation is the microscopic origin of the
Rashba-induced charge visibility discussed below.
The explicit dimensionless Fermi-contour equation, the Rashba branch condition, and the spin-texture components are derived in the Supplemental Material \cite{SupplementalMaterial}.  These formulas are used to generate Fig.~\ref{fig:band_spin_texture}.  The key point is that Rashba coupling does not simply distort the Fermi contours; it also changes the spinor overlaps that enter the response function.  This spinor deformation is the microscopic origin of Rashba brightening.

  More explicitly, the pristine limit has two kinds of structure that must be kept separate.  The Fermi contours are spin resolved, but the spin quantization axis is fixed.  Rashba coupling changes both features at once: it deforms the contours and rotates the spin axis in momentum space.  The contour deformation controls which particle-hole transitions are available, whereas the spin rotation controls the coherence factors of those transitions.  The latter is what allows a density operator to acquire overlap with a longitudinal spin excitation.

\emph{Charge-spin response and collective poles.---}
The gate-screened Coulomb interaction is
\begin{equation}
\vgt=\frac{e^2}{2\epsilon_0\epsilon_r q}\left(1-e^{-2qd_g}\right).
\label{eq:gate_main}
\end{equation}
For $qd_g\ll1$ the gate cuts off the long-range $1/q$ tail, whereas for $qd_g\gg1$ the bare two-dimensional interaction is recovered.  The gate therefore changes the Coulomb feedback but not the single-particle spin texture.

Because Rashba coupling makes spin nonconserved, the response must be formulated in the charge-spin basis $a,b\in(n,S_x,S_y,S_z)$.  With vertices $\Gamma_n=\mathbb{I}$ and $\Gamma_i=\sigma_i$, the band projectors entering the response are
\begin{equation}
P_{\lambda\vk}
\equiv
P_\lambda(\vk)
=
\frac{1}{2}
\left[
\mathbb{I}
+
\lambda \hat{\mathbf d}(\vk)\cdot\boldsymbol\sigma
\right],
\qquad
\lambda=\pm1 .
\label{eq:projector_main}
\end{equation}
The bare response is then the generalized Kubo/Lindhard function \cite{Kubo1957,Lindhard1954,Mahan2000}
\begin{equation}
\begin{aligned}
\chi^{(0)}_{ab}(\vq,\omega)
=&
\sum_{\lambda\lambda'}
\int
\frac{d^2k}{(2\pi)^2}\,
\frac{
f_{\lambda\vk}-f_{\lambda'\vk+\vq}
}{
\hbar\omega+E_{\lambda\vk}-E_{\lambda'\vk+\vq}+i\Gamma
}\\
&\times
\mathrm{Tr}
\!\left[
P_{\lambda\vk}\Gamma_a
P_{\lambda'\vk+\vq}\Gamma_b
\right].
\end{aligned}
\label{eq:chi0main}
\end{equation}
The projector trace is the coherence factor.  It is gauge independent and contains both the ordinary Fermi-surface phase space and the relative rotation of spin texture between $\vk$ and $\vk+\vq$.  In the Supplemental Material we show explicitly that Eq.~\eqref{eq:chi0main} reduces to the spin-resolved Lindhard response in the $\alpha_R\rightarrow0$ limit and yields finite mixed charge-spin components when Rashba spinors are present \cite{SupplementalMaterial}.

The Coulomb interaction couples only to charge density.  The RPA Dyson equation therefore reduces to
\begin{equation}
\chi^{\rm RPA}_{ab}
=
\chi^{(0)}_{ab}
+
\frac{\vgt\chi^{(0)}_{an}\chi^{(0)}_{nb}}
{1-\vgt\chi^{(0)}_{nn}}.
\label{eq:rpamain}
\end{equation}
Equation~\eqref{eq:rpamain} is the mathematical core of the mechanism.  The collective pole is governed by the density-like denominator $1-\vgt\chi_{nn}^{(0)}$, but the spectral residue in any channel is controlled by the numerator.  A mode can therefore remain strongest in $\ASz=-\mathrm{Im}\,\chi^{\rm RPA}_{zz}$ while acquiring a finite density footprint in $\An=-\mathrm{Im}\,\chi^{\rm RPA}_{nn}$.  We quantify this footprint and the internal spin character by
\begin{equation}
\begin{aligned}
\Vch
&=
\frac{\An(q,\omega_d)}{\ASz(q,\omega_d)},\\
W_i
&=
\frac{A_{S_i}(q,\omega_d)}
{A_{S_x}(q,\omega_d)+A_{S_y}(q,\omega_d)+A_{S_z}(q,\omega_d)}.
\end{aligned}
\label{eq:diagnostics_main}
\end{equation}
Here $\omega_d$ is the tracked acoustic spin-demon frequency.  We track the acoustic branch continuously from the pristine spin-demon limit and evaluate both its spectral sharpness and spin character at the same pole; the details of this procedure are given in the Supplemental Material~\cite{SupplementalMaterial}.

The emergence of the Rashba spin demon can already be anticipated at the bare-response level.  Rashba spinor mixing gives the density and longitudinal-spin vertices a finite overlap in the same low-energy window where the acoustic demon branch is formed; the corresponding continuum analysis is given in the Supplemental Material~\cite{SupplementalMaterial}.  These plots establish that Rashba coupling produces mixed charge-spin spectral weight in the same region where the pristine spin demon lives.  Separately, the gate-only analysis in the Supplemental Material shows how electrostatic screening tunes the acoustic branch and quality factor even when $\alpha_R=0$ \cite{SupplementalMaterial}.  Thus Rashba modifies the residues, while the gate modifies the collective denominator.

  In the spin-conserving benchmark, the anisotropic spin-$\sigma$ band can be mapped to an isotropic two-dimensional Lindhard problem by using the projected momentum $\bar q_\sigma=\eta_\sigma(\theta)q/k_F$.  The spin-resolved continuum edges are then separated whenever $\eta_\uparrow\neq\eta_\downarrow$.  In the long-wavelength limit the pristine acoustic spin-demon branch is
\begin{equation}
x_d^{(0)}
=
\frac{\hbar\omega_d^{(0)}}{E_F}
=
\frac{4}{\sqrt{3}}\eta_{\min}\frac{q}{k_F},
\label{eq:pristine_branch_main}
\end{equation}
with $\eta_{\min}=\min(\eta_\uparrow,\eta_\downarrow)$.  We use this branch only as a reference guide in the figures; the full Rashba-plus-gate pole is extracted from the interacting spectral function.  The derivation of Eq.~\eqref{eq:pristine_branch_main}, the continuum edges, and the gate-screened analytical limit is given in the Supplemental Material \cite{SupplementalMaterial}.

\begin{figure}[!htbp]
\centering
\includegraphics[width=0.92\columnwidth]{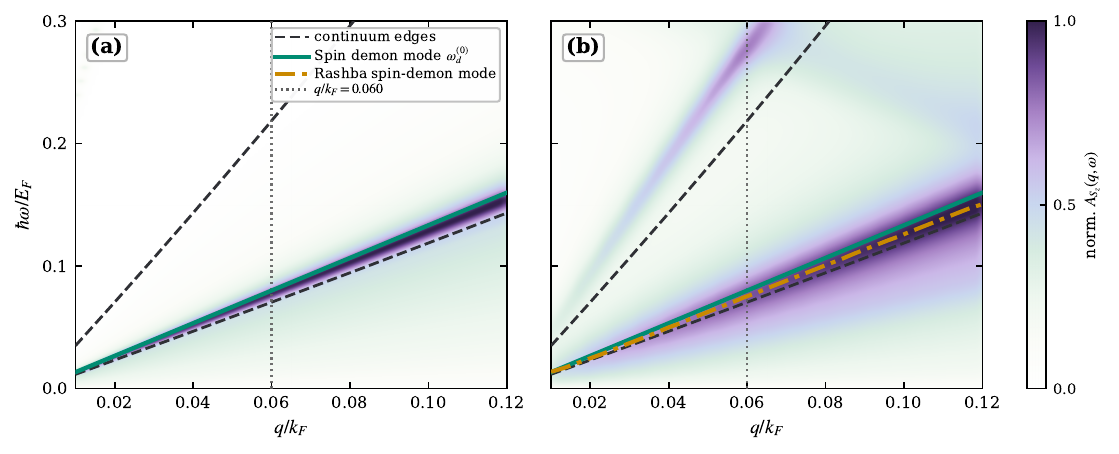}
\vspace{-0.45em}
\caption{Survival of the interacting spin demon.  (a) Pristine longitudinal spin spectral function $\ASz(q,\omega)$ in the spin-conserving limit.  (b) Full Rashba-plus-gate result at $\bar\alpha_R=0.25$ and $k_Fd_g=1.5$.  The mode remains acoustic and continuously connected to the pristine spin demon, while Rashba coupling introduces a spin-mixed background.  The color scales are normalized separately for visual clarity.}
\label{fig:rpa_survival}
\end{figure}

\emph{Survival of the Rashba spin demon.---}
Figure~\ref{fig:rpa_survival} illustrates whether the acoustic spin demon survives once Rashba mixing and gate screening are included.  In the pristine limit, the projected spin-resolved continua have different edges along a principal axis.  The resulting low-damping interval supports an acoustic pole in the longitudinal spin response.  Figure~\ref{fig:rpa_survival}(a) shows this benchmark: the mode follows the pristine reference branch but it is distinct from it and lies in the expected low-energy window.

Figure~\ref{fig:rpa_survival}(b) shows the full Rashba-plus-gate spectrum.  Rashba coupling creates additional background because the spinors are now momentum dependent and the response is no longer diagonal in pure spin.  Nevertheless, the low-energy branch remains visible and tracks the acoustic spin-demon mode.  This continuity is essential: the surviving pole is not a new high-energy charge mode, but the Rashba-dressed version of the pristine altermagnetic spin demon.  The finite gate distance changes the Coulomb feedback, but it does not close the low-energy spin-demon window for the parameters shown.

  The two panels should be read as a continuity test rather than as two unrelated spectra.  Panel (a) fixes the reference problem: the strong $A_{S_z}$ Rashba mode is the spin-conserving spin demon.  Panel (b) asks whether this Rashba mode remains when the spinor basis is changed and the Coulomb interaction is screened.  The answer is positive.  The Rashba mode broadens and the background changes, as expected from Rashba-induced spin-charge mixing, but the low-energy acoustic branch remains visible.  This is the main reason we can meaningfully discuss a ``brightened'' spin demon rather than a newly created charge mode.

\begin{figure}[!htbp]
\centering
\includegraphics[width=0.92\columnwidth]{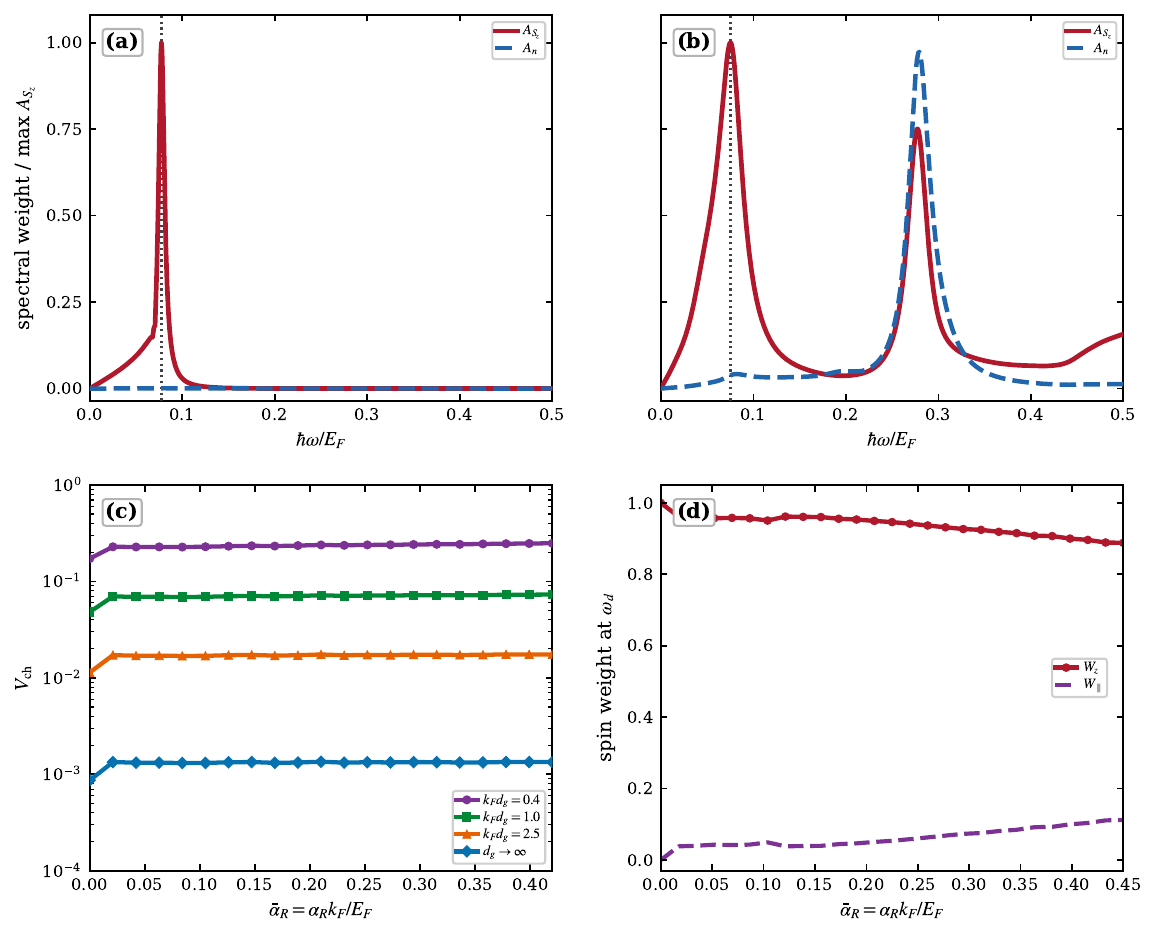}
\vspace{-0.45em}
\caption{Rashba brightening without loss of spin-demon character.  (a) Pristine line cut at $q/k_F=0.060$: the spin demon is nearly charge dark.  (b) Rashba-plus-gate line cut at $\bar\alpha_R=0.25$ and $k_Fd_g=1.5$: the low-energy spin-demon pole acquires finite charge weight while a separate high-energy charge-active feature appears at larger frequency.  (c) Charge visibility $\Vch$ versus Rashba strength for several gate distances.  (d) Spin weights at the extracted spin-demon pole.  The longitudinal component $W_z$ remains dominant, while $W_\parallel=W_x+W_y$ stays subleading.}
\label{fig:charge_visibility}
\end{figure}

\emph{Charge brightening and spin identity.---}
The next question is whether the surviving spin demon becomes visible in the charge response.  Figure~\ref{fig:charge_visibility}(a) shows the pristine line cut at $q/k_F=0.060$.  The peak is strong in $\ASz$ but nearly absent in $\An$, demonstrating the charge-dark character of the out-of-phase spin-density oscillation.  Figure~\ref{fig:charge_visibility}(b) shows the Rashba-plus-gate case at the same momentum.  The low-energy acoustic pole now has a finite charge signal.  The same extended frequency window also contains a higher-energy charge-dominant feature.  This feature is not used to define $\Vch$; the visibility is evaluated at the tracked low-energy spin-demon pole.

Figure~\ref{fig:charge_visibility}(c) shows the visibility ratio as a function of Rashba strength for several gate distances.  The increase with $\bar\alpha_R$ reflects the strengthening of spinor-induced charge-spin mixing.  The dependence on $k_Fd_g$ reflects the separate role of electrostatic screening: changing the gate distance changes how strongly the density channel feeds back through the RPA denominator.  The extended two-parameter maps in the Supplemental Material show that this behavior is a continuous control landscape rather than a single fine-tuned point \cite{SupplementalMaterial}.

  The behavior in Fig.~\ref{fig:charge_visibility}(c) also clarifies the role of the gate.  At fixed Rashba strength, reducing the gate distance screens the long-range interaction and changes how the density component feeds back into the collective response.  This is not equivalent to changing $\bar\alpha_R$: Rashba controls the spinor matrix elements, while the gate controls the Coulomb kernel.  The two knobs are therefore complementary.  In practice, one can use Rashba coupling to create a charge handle and the gate to tune the balance between visibility and damping.

Finally, Fig.~\ref{fig:charge_visibility}(d) shows that brightening does not destroy the spin-demon identity.  The longitudinal weight $W_z$ remains close to unity over the Rashba range shown, whereas $W_\parallel=W_x+W_y$ stays subleading.  The full spin decomposition and representative spin spectral cuts in the Supplemental Material show the same conclusion in more detail \cite{SupplementalMaterial}.  Rashba coupling generates transverse admixture and a finite charge residue, but the low-energy pole remains a longitudinal spin-dominated acoustic mode.

\emph{Physical interpretation and outlook.---}
The results separate two roles that are often entangled in spin-charge collective modes.  Rashba coupling acts at the level of quasiparticle spinors and response-function residues: it gives the spin-demon pole charge visibility.  Gate screening acts at the level of the Coulomb denominator: it tunes the pole position and quality factor.  This is why the mode in Fig.~\ref{fig:charge_visibility} is neither a conventional charge plasmon nor the completely dark pristine spin demon.  It remains distinct from the pristine spin demon branch, but it acquires a finite charge handle.

This distinction is important experimentally.  A purely charge-dark spin demon is naturally suited to spin-sensitive probes, whereas a finite $\An(q,\omega_d)$ opens a route to density-sensitive spectroscopy.  The visibility ratio $\Vch$ is therefore more than a formal quantity: it measures how much of a spin-dominated collective excitation leaks into the experimentally accessible charge channel.  Candidate systems should combine a sizable $d$-wave altermagnetic splitting, a pristine two-dimensional conducting channel, inversion-breaking Rashba coupling, and a controllable dielectric environment.  Material-specific band structures, disorder, phonons, domain structure, and additional spin-orbit terms will affect the quantitative linewidths and visibility, but they do not alter the main results.

  This distinction suggests a concrete experimental reading of the spectra.  A purely charge probe should see very little spectral weight at the pristine spin-demon frequency.  After Rashba brightening, the same probe can acquire a finite response at the acoustic pole, while a spin-sensitive probe should still see the stronger longitudinal spin peak.  Observing both features in the same low-energy window would be a clear signature of a charge-visible spin demon rather than an ordinary plasmon.  The gate dependence provides an additional check, because moving the gate should tune the collective pole without changing the underlying $d$-wave altermagnetic spin splitting.

\emph{Conclusions.---}
We have identified a mode, Rashba spin demon mode, which is distinct from the pristine spin demon mode and is long lived in Rashba 2D $d$-wave altermagnets. It is an acoustic mode which is predominantly a spin mode with a charge character. We have also shown that Rashba coupling and electrostatic gate screening provide complementary controls of spin demons in two-dimensional $d$-wave altermagnets.  The pristine spin demon originates from separated spin-resolved continua and is nearly charge dark.  Rashba spin-orbit coupling converts the pure-spin problem into a charge-spin response problem, producing a finite density residue at the acoustic spin-demon pole.  Gate screening tunes the Coulomb denominator and hence the dispersion and quality factor.  The resulting excitation remains predominantly longitudinal-spin-like while becoming charge visible.  This establishes Rashba-brightened spin demons as tunable spin-charge collective modes in altermagnetic heterostructures.

\bibliography{references}

\onecolumngrid
\clearpage
\setcounter{section}{0}
\setcounter{equation}{0}
\setcounter{figure}{0}
\renewcommand{\thesection}{\Roman{section}}
\renewcommand{\theequation}{S\arabic{equation}}
\renewcommand{\thefigure}{S\arabic{figure}}
\setcounter{secnumdepth}{2}

\section*{Supplementary Material for "Rashba Spin Demons in Two-Dimensional $d$-Wave Altermagnets and their Electrostatic Control"}

This Supplemental Material gives the derivations and numerical definitions underlying the paper. The presentation follows the logical order of the calculation. Sec.~\ref{sec:single_particle} introduces the single-particle model, reference scales, and Rashba-modified Fermi geometry. Sec.~\ref{sec:response_rpa} derives the charge-spin response matrix and the RPA resummation. Sec.~\ref{sec:continua} gives the projected continuum edges and the acoustic guide branch used to track the Rashba spin demon. Sec.~\ref{sec:gate_screening} discusses the gate-screened interaction and the gate-only analytical limit. Sec.~\ref{sec:pole_diagnostics} defines the pole extraction, linewidth, charge visibility, and spin-character diagnostics. Sec.~\ref{sec:maps_checks} summarizes the Rashba-gate maps and the limiting-case checks.

\section{Single-particle structure and dimensionless conventions}
\label{sec:single_particle}

\subsection*{A. Spin-conserving $d$-wave altermagnet}

The pristine continuum Hamiltonian is
\begin{equation}
\begin{aligned}
H_0(\vk)&=\epsilon_0(\vk)\mathbb{I}+\Delta_d(\vk)\sigma_z,\\
\epsilon_0(\vk)&=\frac{\hbar^2k^2}{2m_0},
\qquad
\Delta_d(\vk)=\frac{\hbar^2}{2m^\ast}(k_x^2-k_y^2).
\end{aligned}
\label{eq:S1_H0}
\end{equation}
The $d$-wave term satisfies $\Delta_d\propto k^2\cos2\phi$, changes sign under a $\pi/2$ rotation, and vanishes on the nodal lines $k_x=\pm k_y$.  This is the continuum representation of the nonrelativistic altermagnetic spin splitting used in the paper \cite{Smejkal2022PRX,Smejkal2022Landscape,Gunnink2025PRL}.  Since $[H_0,\sigma_z]=0$, the pristine eigenstates have a conserved spin index $\sigma=\pm1$ and energies
\begin{equation}
E_\sigma(\vk)=
\frac{\hbar^2k^2}{2m_0}
+
\sigma\frac{\hbar^2}{2m^\ast}(k_x^2-k_y^2).
\label{eq:S1_Esigma}
\end{equation}
Collecting the coefficients of $k_x^2$ and $k_y^2$ gives the anisotropic effective-mass form
\begin{equation}
\begin{aligned}
E_\sigma(\vk)
&=
\frac{\hbar^2 k_x^2}{2m_x^\sigma}
+
\frac{\hbar^2 k_y^2}{2m_y^\sigma},\\
m_x^\sigma&=\frac{m_0m^\ast}{m^\ast+\sigma m_0},
\qquad
m_y^\sigma=\frac{m_0m^\ast}{m^\ast-\sigma m_0}.
\end{aligned}
\label{eq:S1_masses}
\end{equation}
The two spin contours are rotated relative to one another because
\begin{equation}
m_x^\uparrow=m_y^\downarrow,
\qquad
m_y^\uparrow=m_x^\downarrow.
\label{eq:S1_rotated_masses}
\end{equation}
This rotated-ellipse geometry is the phase-space origin of the altermagnetic spin-demon window: a perturbation with fixed direction samples different projected momenta in the two spin channels.

The density-of-states mass is independent of $\sigma$,
\begin{equation}
m_{\rm DOS}
=
\sqrt{m_x^\sigma m_y^\sigma}
=
\frac{m_0m^\ast}{\sqrt{(m^\ast)^2-m_0^2}},
\label{eq:S1_mdos}
\end{equation}
and defines the reference scales
\begin{equation}
\begin{aligned}
k_F&=\frac{\sqrt{2m_{\rm DOS}E_F}}{\hbar},
&
 v_F&=\frac{\hbar k_F}{m_{\rm DOS}},
&
 N_0&=\frac{m_{\rm DOS}}{2\pi\hbar^2}.
\end{aligned}
\label{eq:S1_scales}
\end{equation}
With $\kappa_i=k_i/k_F$, $A_0=m_{\rm DOS}/m_0$, and $A_d=m_{\rm DOS}/m^\ast$, Eq.~\eqref{eq:S1_Esigma} becomes
\begin{equation}
\frac{E_\sigma}{E_F}
=
A_0(\kappa_x^2+\kappa_y^2)
+
\sigma A_d(\kappa_x^2-\kappa_y^2).
\label{eq:S1_dimless_pristine}
\end{equation}
For $\kappa_x=\kappa\cos\phi$ and $\kappa_y=\kappa\sin\phi$, the Fermi contour $E_\sigma=E_F$ is
\begin{equation}
\kappa_\sigma(\phi)
=
\left[A_0+\sigma A_d\cos2\phi\right]^{-1/2}.
\label{eq:S1_pristine_contour}
\end{equation}
Equation~\eqref{eq:S1_pristine_contour} gives the spin-resolved contours plotted in Fig.~2(a) of the paper.

\subsection*{B. Rashba-modified bands and spin texture}

Interfacial inversion breaking adds the Rashba term $\alpha_R(k_y\sigma_x-k_x\sigma_y)$ \cite{BychkovRashba1984,Nitta1997,Manchon2015,Bihlmayer2022}.  The full Hamiltonian can be written as
\begin{equation}
\begin{aligned}
H(\vk)
&=
\epsilon_0(\vk)\mathbb{I}
+
\Delta_d(\vk)\sigma_z
+
\alpha_R(k_y\sigma_x-k_x\sigma_y)\\
&=
\epsilon_0(\vk)\mathbb{I}
+
\mathbf d(\vk)\cdot\boldsymbol\sigma,
\qquad
\mathbf d(\vk)=(\alpha_R k_y,-\alpha_R k_x,\Delta_d).
\end{aligned}
\label{eq:S1_Hrashba}
\end{equation}
The magnitude of the effective field is
\begin{equation}
|\mathbf d(\vk)|=
\sqrt{\Delta_d^2(\vk)+\alpha_R^2 k^2}.
\label{eq:S1_d_magnitude}
\end{equation}
The eigenvalues and projectors are
\begin{equation}
\begin{aligned}
E_\lambda(\vk)&=\epsilon_0(\vk)+\lambda |\mathbf d(\vk)|,\\
P_\lambda(\vk)&=\frac{1}{2}
\left[
\mathbb{I}
+
\lambda\hat{\mathbf d}(\vk)\cdot\boldsymbol\sigma
\right],
\qquad \lambda=\pm1.
\end{aligned}
\label{eq:S1_bands_projectors}
\end{equation}
Using $\bar\alpha_R=\alpha_Rk_F/E_F$, the dimensionless Rashba-altermagnetic dispersion is
\begin{equation}
\frac{E_\lambda(\kappa,\phi)}{E_F}
=
A_0\kappa^2
+
\lambda
\sqrt{
A_d^2\kappa^4\cos^22\phi
+
\bar\alpha_R^2\kappa^2
}.
\label{eq:S1_dimless_rashba}
\end{equation}
The Fermi contours satisfy $E_\lambda=E_F$.  Setting $y=\kappa^2$, one obtains
\begin{equation}
\begin{aligned}
&\left(A_0^2-A_d^2\cos^22\phi\right)y^2
-
\left(2A_0+\bar\alpha_R^2\right)y
+1=0,\\
&\lambda(1-A_0y)>0.
\end{aligned}
\label{eq:S1_rashba_contour}
\end{equation}
The second line is the branch condition inherited from the unsquared equation.  It fixes which solution belongs to which Rashba-altermagnetic band and removes spurious roots.

The spin expectation value in band $\lambda$ is
\begin{equation}
\begin{aligned}
\langle\boldsymbol\sigma\rangle_\lambda
&=\lambda\hat{\mathbf d}(\vk),\\
\langle\sigma_x\rangle_\lambda
&=\lambda\frac{\bar\alpha_R\kappa\sin\phi}{D(\kappa,\phi)},
&
\langle\sigma_y\rangle_\lambda
&=-\lambda\frac{\bar\alpha_R\kappa\cos\phi}{D(\kappa,\phi)},
&
\langle\sigma_z\rangle_\lambda
&=\lambda\frac{A_d\kappa^2\cos2\phi}{D(\kappa,\phi)},
\end{aligned}
\label{eq:S1_spin_texture}
\end{equation}
where
\begin{equation}
D(\kappa,\phi)
=
\sqrt{A_d^2\kappa^4\cos^22\phi+\bar\alpha_R^2\kappa^2}.
\label{eq:S1_D}
\end{equation}
Thus Rashba coupling reshapes the Fermi contours and rotates the quasiparticle spin texture away from a fixed $\sigma_z$ axis.  The spin texture, rather than the energy shift alone, is what produces mixed charge-spin response in the next section.

\section{Charge-spin response and RPA resummation}
\label{sec:response_rpa}

\subsection*{A. Kubo response in the Rashba band basis}

For $\alpha_R\neq0$, spin is not a conserved quantum number.  We therefore use the charge-spin vertices
\begin{equation}
\Gamma_n=\mathbb{I},
\qquad
\Gamma_i=\sigma_i,
\qquad i=x,y,z.
\label{eq:S2_vertices}
\end{equation}
The generalized density operator and retarded response are
\begin{equation}
\begin{aligned}
\rho_a(\vq)&=\sum_{\vk}c^\dagger_{\vk+\vq}\Gamma_a c_{\vk},\\
\chi_{ab}(\vq,t)&=-\frac{i}{\hbar}\Theta(t)
\left\langle
[\rho_a(\vq,t),\rho_b(-\vq,0)]
\right\rangle .
\end{aligned}
\label{eq:S2_density_response}
\end{equation}
The bare response in the band basis is the generalized Kubo/Lindhard function \cite{Kubo1957,Lindhard1954,Mahan2000},
\begin{equation}
\begin{aligned}
\chi^{(0)}_{ab}(\vq,\omega)
=&
\sum_{\lambda\lambda'}
\int\frac{d^2k}{(2\pi)^2}
\frac{
 f[E_\lambda(\vk)]-f[E_{\lambda'}(\vk+\vq)]
}{
 \hbar\omega+E_\lambda(\vk)-E_{\lambda'}(\vk+\vq)+i\Gamma
}
\,M^{ab}_{\lambda\lambda'}(\vk,\vq),\\
M^{ab}_{\lambda\lambda'}(\vk,\vq)
=&
\mathrm{Tr}
\left[
P_\lambda(\vk)\Gamma_a
P_{\lambda'}(\vk+\vq)\Gamma_b
\right].
\end{aligned}
\label{eq:S2_chi0}
\end{equation}
The trace factor is gauge independent and keeps the spinor overlaps explicitly.

Let
\begin{equation}
\mathbf n=\hat{\mathbf d}(\vk),
\qquad
\mathbf n'=\hat{\mathbf d}(\vk+\vq).
\label{eq:S2_nvectors}
\end{equation}
Using the Pauli identities
\begin{equation}
\begin{aligned}
\mathrm{Tr}[\sigma_i\sigma_j]&=2\delta_{ij},
&
\mathrm{Tr}[\sigma_i\sigma_j\sigma_k]&=2i\epsilon_{ijk},\\
\mathrm{Tr}[\sigma_i\sigma_j\sigma_k\sigma_l]
&=
2(\delta_{ij}\delta_{kl}-\delta_{ik}\delta_{jl}+\delta_{il}\delta_{jk}),
\end{aligned}
\label{eq:S2_pauli_traces}
\end{equation}
one obtains
\begin{equation}
\begin{aligned}
M_{\lambda\lambda'}^{nn}
&=
\frac{1}{2}
\left(1+\lambda\lambda'\mathbf n\cdot\mathbf n'\right),\\
M_{\lambda\lambda'}^{ni}
&=
\frac{1}{2}
\left[
\lambda n_i+\lambda'n_i'
+i\lambda\lambda'(\mathbf n\times\mathbf n')_i
\right],\\
M_{\lambda\lambda'}^{in}
&=
\frac{1}{2}
\left[
\lambda n_i+\lambda'n_i'
-i\lambda\lambda'(\mathbf n\times\mathbf n')_i
\right],\\
M_{\lambda\lambda'}^{ij}
&=
\frac{1}{2}
\left[
\delta_{ij}
+
\lambda\lambda'
(n_i n'_j+n_j n'_i-\delta_{ij}\mathbf n\cdot\mathbf n')
+
i\epsilon_{ij\ell}(\lambda n_\ell-\lambda'n'_\ell)
\right].
\end{aligned}
\label{eq:S2_coherence_factors}
\end{equation}
For the longitudinal spin channel,
\begin{equation}
M_{\lambda\lambda'}^{zz}
=
\frac{1}{2}
\left[
1+
\lambda\lambda'
(2n_zn'_z-\mathbf n\cdot\mathbf n')
\right].
\label{eq:S2_Mzz}
\end{equation}
The mixed factors $M^{ni}$ and $M^{in}$ contain the essential Rashba information.  Their real parts depend on the spin projection carried by the two states, while the imaginary parts are controlled by the relative rotation of the spin texture between $\vk$ and $\vk+\vq$.  These terms are absent in a globally spin-collinear description and are responsible for the density residue of the Rashba spin demon.

\subsection*{B. RPA response and the spin-conserving check}

The Coulomb interaction is density-density only,
\begin{equation}
V_{ab}(\vq)=v_q\delta_{an}\delta_{bn}.
\label{eq:S2_Vab}
\end{equation}
The RPA equation is
\begin{equation}
\chi^{\rm RPA}
=
\chi^{(0)}+\chi^{(0)}V\chi^{\rm RPA}.
\label{eq:S2_RPA_matrix}
\end{equation}
Since $V$ has only one nonzero component,
\begin{equation}
\chi^{\rm RPA}_{ab}
=
\chi^{(0)}_{ab}
+
v_q\chi^{(0)}_{an}\chi^{\rm RPA}_{nb}.
\label{eq:S2_RPA_step}
\end{equation}
Setting $a=n$ gives
\begin{equation}
\chi^{\rm RPA}_{nb}
=
\frac{\chi^{(0)}_{nb}}{1-v_q\chi^{(0)}_{nn}},
\label{eq:S2_RPA_nb}
\end{equation}
and therefore
\begin{equation}
\chi^{\rm RPA}_{ab}
=
\chi^{(0)}_{ab}
+
\frac{
v_q\chi^{(0)}_{an}\chi^{(0)}_{nb}
}{
1-v_q\chi^{(0)}_{nn}
}.
\label{eq:S2_RPA_final}
\end{equation}
The two responses used most directly in the paper are
\begin{equation}
\begin{aligned}
\chi^{\rm RPA}_{nn}
&=
\frac{\chi^{(0)}_{nn}}{1-v_q\chi^{(0)}_{nn}},\\
\chi^{\rm RPA}_{zz}
&=
\chi^{(0)}_{zz}
+
\frac{
v_q\chi^{(0)}_{zn}\chi^{(0)}_{nz}
}{
1-v_q\chi^{(0)}_{nn}
}.
\end{aligned}
\label{eq:S2_RPA_nn_zz}
\end{equation}
Thus the pole denominator is common, but the residue is channel dependent.  This is the formal reason why the same collective pole can be spin dominated and still acquire a finite charge signal.

The spin-conserving limit verifies the construction.  Define
\begin{equation}
S=\chi_\uparrow^{(0)}+\chi_\downarrow^{(0)},
\qquad
D=\chi_\uparrow^{(0)}-\chi_\downarrow^{(0)}.
\label{eq:S2_S_D}
\end{equation}
Then
\begin{equation}
\chi_{nn}^{(0)}=\chi_{zz}^{(0)}=S,
\qquad
\chi_{nz}^{(0)}=\chi_{zn}^{(0)}=D.
\label{eq:S2_pristine_matrix}
\end{equation}
Substitution into Eq.~\eqref{eq:S2_RPA_nn_zz} yields
\begin{equation}
\chi^{\rm RPA}_{zz}
=
S+\frac{v_qD^2}{1-v_qS}
=
\frac{
S-4v_q\chi_\uparrow^{(0)}\chi_\downarrow^{(0)}
}{
1-v_qS
},
\label{eq:S2_pristine_RPA}
\end{equation}
where $D^2-S^2=-4\chi_\uparrow^{(0)}\chi_\downarrow^{(0)}$ has been used.  Equation~\eqref{eq:S2_pristine_RPA} is the spin-resolved RPA form of the pristine altermagnetic spin-demon response \cite{Gunnink2025PRL,Rostami2025FermiLiquid}.

\section{Spin-resolved continua and acoustic reference branch}
\label{sec:continua}

For a perturbation momentum $\vq=q(\cos\theta,\sin\theta)$, the anisotropic spin-$\sigma$ band maps to an isotropic two-dimensional Lindhard problem with projected momentum
\begin{equation}
\begin{aligned}
\bar q_\sigma&=\bar q\,\eta_\sigma(\theta),
\qquad
\bar q=\frac{q}{k_F},\\
\eta_\sigma(\theta)
&=
\left[
\frac{m_{\rm DOS}}{m_x^\sigma}\cos^2\theta
+
\frac{m_{\rm DOS}}{m_y^\sigma}\sin^2\theta
\right]^{1/2}.
\end{aligned}
\label{eq:S3_projected_q}
\end{equation}
The dimensionless frequency is $x=\hbar\omega/E_F$, and we define
\begin{equation}
\nu_{\sigma,\mp}
=
\frac{x}{2\bar q_\sigma}
\mp
\frac{\bar q_\sigma}{2}.
\label{eq:S3_nu}
\end{equation}
The zero-temperature two-dimensional Lindhard function is \cite{Stern1967,GiulianiVignale2005}
\begin{equation}
\frac{\chi_\sigma^{(0)}}{N_0}
=
-
\left(A_\sigma^{2D}+iB_\sigma^{2D}\right),
\label{eq:S3_lindhard}
\end{equation}
with
\begin{equation}
\begin{aligned}
A_\sigma^{2D}
=&
1+
\frac{1}{\bar q_\sigma}
\Big[
\Theta(\nu_{\sigma,-}^2-1)
\operatorname{sgn}(\nu_{\sigma,-})
\sqrt{\nu_{\sigma,-}^2-1}\\
&\hspace{2.5cm}
-
\Theta(\nu_{\sigma,+}^2-1)
\operatorname{sgn}(\nu_{\sigma,+})
\sqrt{\nu_{\sigma,+}^2-1}
\Big],\\
B_\sigma^{2D}
=&
\frac{1}{\bar q_\sigma}
\Big[
\Theta(1-\nu_{\sigma,-}^2)
\sqrt{1-\nu_{\sigma,-}^2}\\
&\hspace{2.5cm}
-
\Theta(1-\nu_{\sigma,+}^2)
\sqrt{1-\nu_{\sigma,+}^2}
\Big].
\end{aligned}
\label{eq:S3_lindhard_AB}
\end{equation}
The particle-hole continuum is the region in which $B_\sigma^{2D}$ is nonzero.  In the small-$\bar q_\sigma$ regime used in the spectra, the upper edge is
\begin{equation}
x_{+,\sigma}
=
\bar q_\sigma^2+2\bar q_\sigma.
\label{eq:S3_continuum_edge}
\end{equation}
When $\eta_\uparrow\neq\eta_\downarrow$, the two spin-resolved continuum edges separate.  This separation opens the low-damping interval that supports the acoustic spin demon.  In the long-wavelength strong-coupling limit, the acoustic reference branch is
\begin{equation}
\begin{aligned}
x_d^{(0)}
&=
\frac{\hbar\omega_d^{(0)}}{E_F}
=
\frac{4}{\sqrt{3}}\eta_{\min}\bar q,
\qquad
\eta_{\min}=\min(\eta_\uparrow,\eta_\downarrow),\\
\frac{\eta_{\min}}{\eta_{\max}}&<\frac{\sqrt{3}}{2},
\qquad
Q_{\rm pristine}
=
\frac{3\sqrt{3\eta_{\max}^2-4\eta_{\min}^2}}{\eta_{\min}},
\qquad
\eta_{\max}=\max(\eta_\uparrow,\eta_\downarrow).
\end{aligned}
\label{eq:S3_branch_Q}
\end{equation}
The first line gives the guide branch.  The second line gives the existence condition and the corresponding analytical quality-factor estimate for the spin-conserving case.  In the Rashba-plus-gate problem, Eq.~\eqref{eq:S3_branch_Q} is used only to identify the acoustic branch; the actual pole is extracted from the full $\chi_{zz}^{\rm RPA}$.

\begin{figure}[t]
\centering
\includegraphics[width=0.82\textwidth]{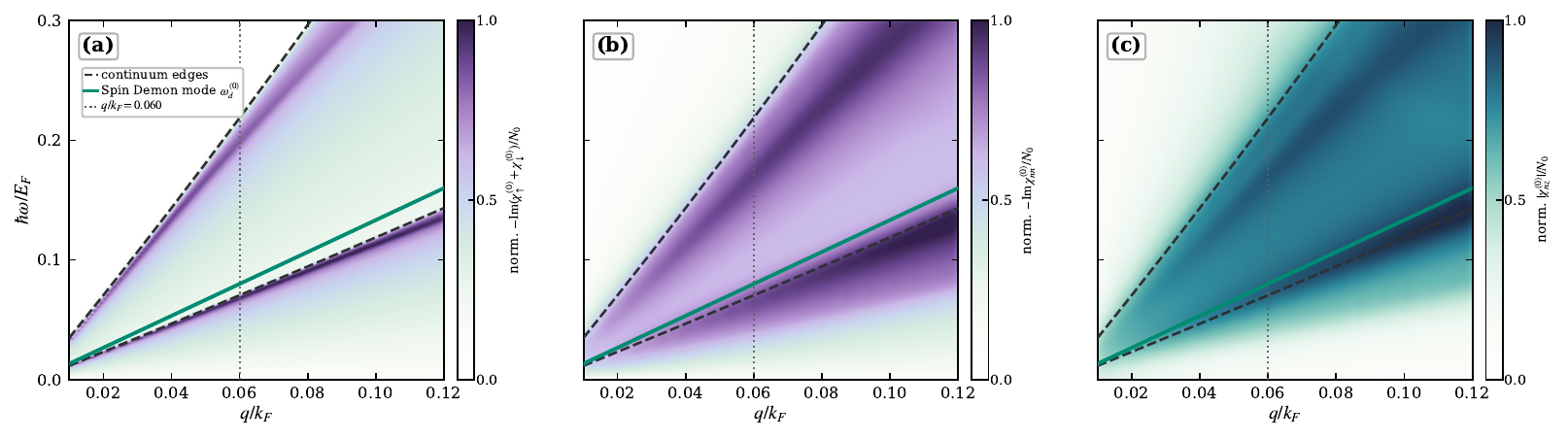}
\caption{Bare continua and Rashba-induced charge-spin mixing in the low-energy acoustic window.  (a) Spin-conserving particle-hole continuum, with dashed curves denoting the spin-resolved continuum edges and the green line giving the acoustic guide branch.  (b) Rashba charge continuum in the same momentum and frequency window.  (c) Mixed charge-spin response $|\chi_{nz}^{(0)}|/N_0$.  The finite mixed response in the spin-demon window identifies the bare channel through which a longitudinal spin excitation acquires density spectral weight after RPA feedback.}
\label{figS:bare_mixing}
\end{figure}

Figure~\ref{figS:bare_mixing} shows the response-level origin of the Rashba spin demon.  The spin-conserving continuum separation defines the acoustic window.  Rashba spinor mixing then gives a finite $\chi_{nz}^{(0)}$ in the same frequency range, providing the channel through which the longitudinal spin pole gains density spectral weight after RPA feedback.

\section{Gate screening and gate-only analytical control}
\label{sec:gate_screening}

A metallic gate at distance $d_g$ produces image-charge screening of the two-dimensional Coulomb interaction \cite{Chaplik1972,SarmaMadhukar1981,Torre2019,Zabolotnykh2019}.  The interaction used in the paper is
\begin{equation}
\begin{aligned}
v_q^{\rm gate}
&=
\frac{e^2}{2\epsilon_0\epsilon_r q}
\left(1-e^{-2qd_g}\right),\\
\frac{v_q^{\rm gate}}{v_q^{(0)}}
&=
1-e^{-2qd_g}.
\end{aligned}
\label{eq:S4_gate_interaction}
\end{equation}
The limiting forms are
\begin{equation}
\begin{aligned}
v_q^{\rm gate}&\rightarrow v_q^{(0)},
& qd_g&\gg1,\\
v_q^{\rm gate}&\simeq \frac{e^2d_g}{\epsilon_0\epsilon_r},
& qd_g&\ll1.
\end{aligned}
\label{eq:S4_gate_limits}
\end{equation}
Thus the gate cuts off the small-$q$ Coulomb singularity without changing the Rashba spin texture.

The dimensionless screened coupling can be written as
\begin{equation}
\alpha_{\rm gate}(\bar q,\bar d_g)
=
v_q^{\rm gate}N_0
=
g_C\frac{1-e^{-2\bar q\bar d_g}}{\bar q},
\qquad
g_C=\frac{e^2N_0}{2\epsilon_0\epsilon_r k_F},
\qquad
\bar d_g=k_Fd_g.
\label{eq:S4_alpha_gate}
\end{equation}
In the spin-conserving gate-only limit, the acoustic pole is
\begin{equation}
\begin{aligned}
x_d^{\rm gate}
&=2\bar q\,u_g\eta_{\min},\\
u_g
&=
\frac{C_g}{\sqrt{C_g^2-1}},
\qquad
C_g=2+\frac{1}{\alpha_{\rm gate}},\\
Q_{\rm gate}
&=
\frac{\sqrt{1-(\delta u_g)^2}}{\delta(u_g^2-1)^{3/2}},
\qquad
\delta=\frac{\eta_{\min}}{\eta_{\max}}.
\end{aligned}
\label{eq:S4_gate_branch_Q}
\end{equation}
For $\alpha_{\rm gate}\rightarrow\infty$, $u_g\rightarrow2/\sqrt3$, and Eq.~\eqref{eq:S4_gate_branch_Q} reduces to the acoustic branch in Eq.~\eqref{eq:S3_branch_Q}.  This limit checks the normalization of the screened interaction.

\begin{figure}[t]
\centering
\includegraphics[width=0.74\textwidth]{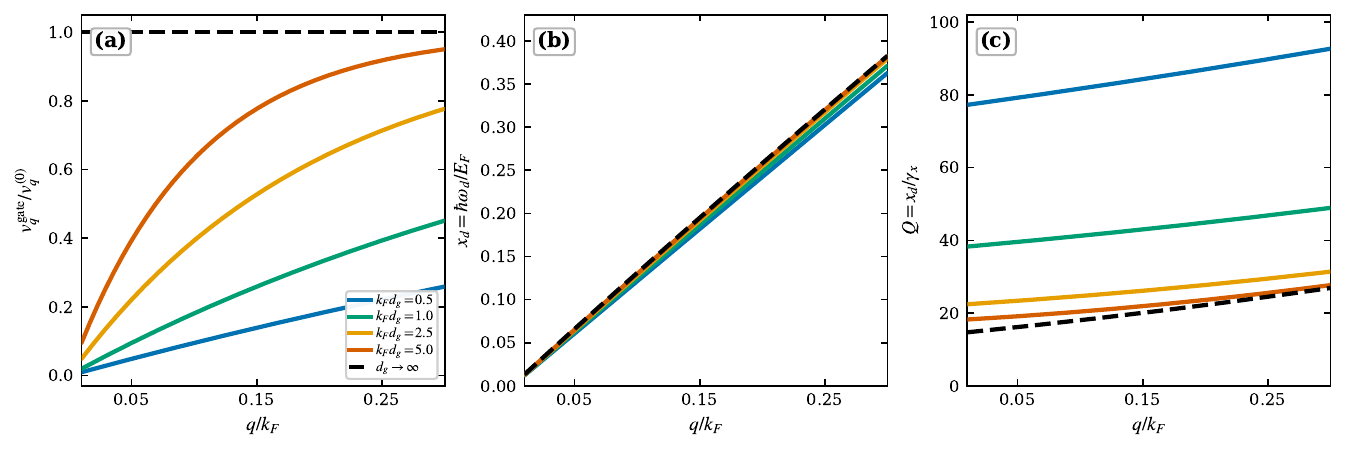}
\caption{Gate-only analytical control of the acoustic spin-demon branch in the spin-conserving limit.  (a) Screened interaction ratio $v_q^{\rm gate}/v_q^{(0)}$ for several gate distances, showing the cutoff of the small-$q$ Coulomb tail.  (b) Gate-controlled dimensionless acoustic frequency $x_d=\hbar\omega_d/E_F$.  (c) Quality factor $Q=x_d/\gamma_x$.  The gate modifies the collective denominator and therefore tunes both the mode velocity and spectral sharpness without changing the single-particle spin texture.}
\label{figS:gate_control}
\end{figure}

Figure~\ref{figS:gate_control} isolates electrostatic control.  Reducing $k_Fd_g$ suppresses the long-range Coulomb feedback, softens the acoustic branch, and changes the quality factor.  This confirms that the gate controls the collective denominator, whereas Rashba coupling controls the response residues through spinor coherence factors.

\section{Pole extraction and spectral diagnostics}
\label{sec:pole_diagnostics}

The Rashba spin demon is identified as the low-energy acoustic pole continuously connected to the spin-conserving altermagnetic spin demon.  The pole is tracked in
\begin{equation}
A_{S_z}(q,\omega)
=
-\mathrm{Im}\,\chi_{zz}^{\rm RPA}(q,\omega).
\label{eq:S5_ASz}
\end{equation}
At fixed $q$, we define
\begin{equation}
\omega_d(q)
=
\arg\max_{\omega\in W_d}A_{S_z}(q,\omega),
\label{eq:S5_pole_tracking}
\end{equation}
where $W_d$ is a low-energy window around the acoustic branch.  The restriction to $W_d$ prevents a higher-energy Rashba-mixed charge feature from being misidentified as the spin demon.

The dielectric function used for the linewidth estimate is
\begin{equation}
\epsilon(q,\omega)
=
1-v_q^{\rm gate}\chi_{nn}^{(0)}(q,\omega).
\label{eq:S5_dielectric}
\end{equation}
For a sharp pole, the spectral maximum agrees with $\mathrm{Re}\,\epsilon(q,\omega_d)=0$.  The linewidth, quality factor, charge visibility, and spin weights are then evaluated at the same pole:
\begin{equation}
\begin{aligned}
\gamma_x
&=
\left|
\frac{\mathrm{Im}\,\epsilon(q,x_d)}{\partial_x\mathrm{Re}\,\epsilon(q,x)|_{x=x_d}}
\right|,
&
Q&=\frac{x_d}{\gamma_x},
&
x_d&=\frac{\hbar\omega_d}{E_F},\\
V_{\rm ch}
&=
\frac{A_n(q,\omega_d)}{A_{S_z}(q,\omega_d)},
&
A_n(q,\omega)&=-\mathrm{Im}\,\chi_{nn}^{\rm RPA}(q,\omega),\\
W_i
&=
\frac{A_{S_i}(q,\omega_d)}{A_{S_x}(q,\omega_d)+A_{S_y}(q,\omega_d)+A_{S_z}(q,\omega_d)},
&
W_\parallel&=W_x+W_y.
\end{aligned}
\label{eq:S5_diagnostics}
\end{equation}
A charge-visible Rashba spin demon is therefore characterized by finite $V_{\rm ch}$ and dominant longitudinal spin weight $W_z$.

\begin{figure}[t]
\centering
\includegraphics[width=0.60\textwidth]{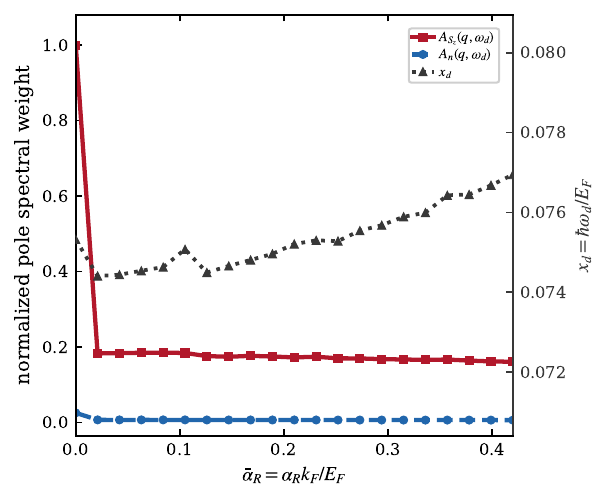}
\caption{Pole spectral weights and extracted acoustic frequency versus Rashba strength at $q/k_F=0.060$ and $k_Fd_g=1.5$.  The longitudinal-spin and charge weights are evaluated at the tracked low-energy pole, while the extracted frequency $x_d=\hbar\omega_d/E_F$ verifies the continuous evolution of the acoustic branch.  The finite but subleading charge weight is the spectral signature of a Rashba-brightened spin demon rather than a conventional plasmon.}
\label{figS:pole_weights}
\end{figure}

Figure~\ref{figS:pole_weights} verifies that the extracted branch evolves smoothly with $\bar\alpha_R$.  The charge pole weight becomes finite but remains below the longitudinal spin weight, which is the signature of a spin-dominated acoustic pole with a finite density residue.

\begin{figure}[t]
\centering
\includegraphics[width=0.72\textwidth]{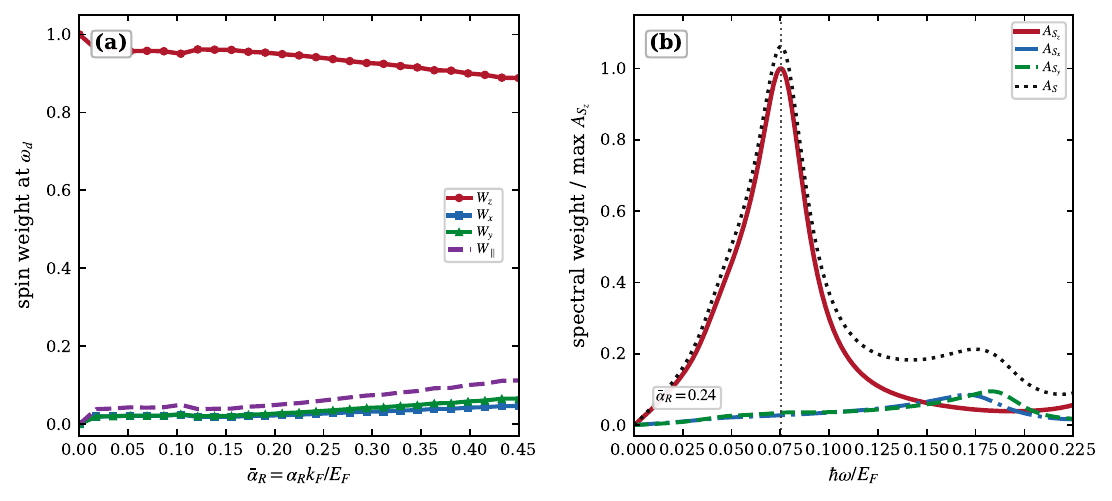}
\caption{Spin-character decomposition of the Rashba-brightened acoustic pole.  (a) Spin weights $W_z$, $W_x$, $W_y$, and $W_\parallel=W_x+W_y$ evaluated at the extracted pole as functions of Rashba strength.  (b) Representative spin spectral cuts at finite Rashba coupling.  The dominant low-energy spectral weight remains in the longitudinal channel $A_{S_z}$, while Rashba coupling produces only subleading transverse admixture.}
\label{figS:spin_character}
\end{figure}

Figure~\ref{figS:spin_character} confirms that the brightened acoustic pole remains mainly longitudinal.  Rashba coupling generates in-plane spin admixture, but the strongest low-energy spectral weight remains in $A_{S_z}$.

\section{Rashba-gate maps and consistency checks}
\label{sec:maps_checks}

\begin{figure}[t]
\centering
\includegraphics[width=0.72\textwidth]{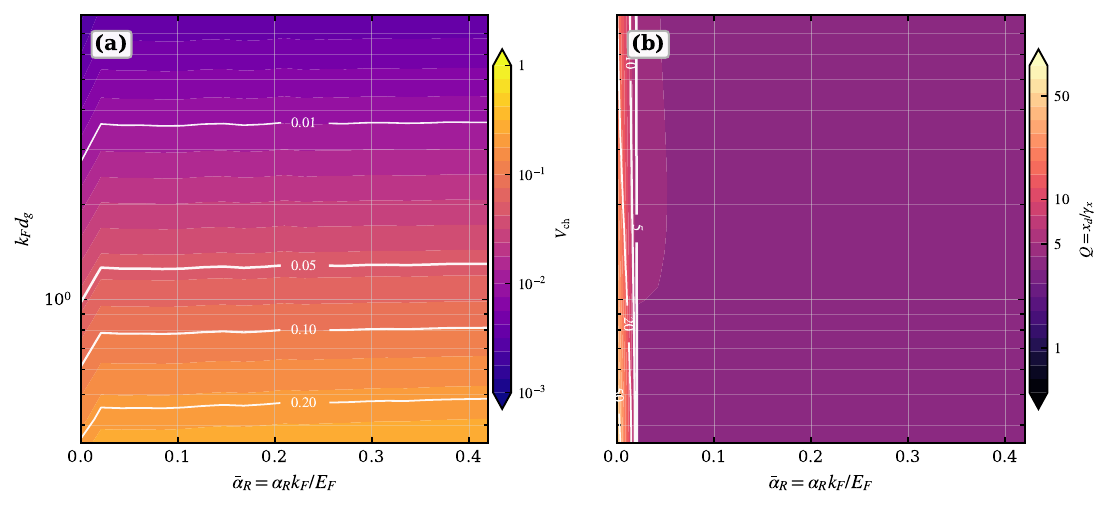}
\caption{Rashba-gate control maps for the acoustic Rashba spin demon.  (a) Charge visibility $V_{\rm ch}$ in the $(\bar\alpha_R,k_Fd_g)$ plane.  (b) Quality factor $Q=x_d/\gamma_x$ in the same control plane.  Horizontal motion changes Rashba spinor mixing and hence the charge-spin residues, whereas vertical motion changes gate screening and hence the collective denominator.  The useful operating regime is where the pole is both charge visible and sufficiently sharp.}
\label{figS:control_maps}
\end{figure}

Figure~\ref{figS:control_maps} displays the two independent external controls.  Varying $\bar\alpha_R$ changes the Rashba spinor texture and hence the response residues.  Varying $k_Fd_g$ changes the screened Coulomb interaction and hence the collective denominator.  The useful regime is where the pole is both charge visible and sufficiently sharp.

The implementation was checked in the following limiting cases.  For $\alpha_R=0$ and $d_g\rightarrow\infty$, the pristine spin-resolved altermagnetic spin-demon response is recovered.  For $\alpha_R=0$ and finite $d_g$, the band structure remains spin conserving and only the Coulomb interaction is screened.  For $\alpha_R\neq0$ and $d_g\rightarrow\infty$, the response contains Rashba coherence factors with the bare Coulomb interaction.  For $\alpha_R\neq0$ and finite $d_g$, the full result is
\begin{equation}
\chi^{\rm RPA}_{ab}
=
\chi^{(0)}_{ab}
+
\frac{
v_q^{\rm gate}\chi^{(0)}_{an}\chi^{(0)}_{nb}
}{
1-v_q^{\rm gate}\chi^{(0)}_{nn}
}.
\label{eq:S6_full_response}
\end{equation}
The numerical code reproduces the pristine contour in Eq.~\eqref{eq:S1_pristine_contour}, the Rashba contour and branch condition in Eq.~\eqref{eq:S1_rashba_contour}, the continuum edge in Eq.~\eqref{eq:S3_continuum_edge}, and the gate-screening ratio in Eq.~\eqref{eq:S4_gate_interaction}.  These checks ensure that the charge-visible pole discussed in the paper is continuously connected to the altermagnetic spin demon.

\end{document}